\documentclass{phbauth}
\usepackage{graphicx}

\begin{document}

\begin{frontmatter}

\title{Thermobaric Effect on Melt-Textured MBa$_2$Cu$_3$O$_{7-\delta}$
(M = Y, Nd)}

\author[address1]{T. Prikhna\thanksref{thank1}},
\author[address2]{W. Gawalek},
\author[address1]{V. Moshchil},
\author[address3]{R. Viznichenko},
\author[address4]{F. Sandiumenge},
\author[address5]{V. Melnikov},
\author[address6]{P. Shaetzle},
\author[address1]{P. Nagorny},
\author[address2]{A. Surzhenko},
\author[address1]{S. Dub},
\author[address2]{Ch. Wende}

\address[address1]{Institute for Superhard Materials, 2,
Avtozavodskaya Str., Kiev 254074 , Ukraine}
\address[address2]{Institut fuer Physikalische Hochtechnologie e.V., 10,
Winzerlaer, Jena D-07745, Germany}
\address[address3]{Institute of Metal Physics, 36, Vernadsky Ave., Kiev,
252142, Ukraine}
\address[address4]{Instituto de Ciencia de Materiales de Barcelona, Bellaterra
E-08193, Catalunya, Espana}
\address[address5]{Institute of Geochemistry, Mineralogy and Ore-Formation,
34, Palladin Ave., Kiev 252142, Ukraine}
\address[address6]{Institut fuer Festkoerper- und Werkstofforschung,
Winterbergstrasse 28, D-01277 Dresden, Germany}

\thanks[thank1]{Corresponding author. Present address: Institute for Superhard
Materials, 2, Avtozavodskaya Str., Kiev 254074 , Ukraine,
E-mail: frd@ismanu.kiev.ua}

\begin{abstract}
The effect of a short (10--30~min) thermobaric action (in the 1--5~GPa pressure 
and 700--1300$^\circ$C temperature range) on the structure, superconductive and 
mechanical properties of melt--textured-- MBa$_2$Cu$_3$O$_{7-\delta}$ (M=Y, Nd) 
or MT--MBCO have been studied. The existence has been established of 
pressure--temperature--time conditions (2~GPa, 800$^\circ$C for 30~min and 
900--950$^\circ$C for 15~min for MT--YBCO; 5~GPa, 850--900$^\circ$C for 15~min 
for MT--NdBCO) the treatment under which allows superconductive properties of 
the materials (because of the contact with zirconia and high pressure) to be 
preserved or improved, mechanical characteristics increased and the materials 
condensed.
\end{abstract}


\end{frontmatter}

\section{Introduction}

Among known bulk high-temperature superconductive materials, MT-MBCO (M=Y, 
Nd) possesses the highest level of cryomagnetic properties, which enables us
to construct HTSC electromotors with massive MT-YBCO rotors with 
output power of about 19 kW at 3000 rpm \cite{1}. The structure of MT-MBCO
samples consists of a few large  textured grains or magnetic domains of
MBa$_2$Cu$_3$O$_{7-\delta}$ (M123) with finely dispersed small 
inclusions of the non-superconductive Y$_2$BaCuO$_5$ (Y211) phase or
Nd$_4$Ba$_2$Cu$_2$O$_{10}$ (Nd422) for M=Y or  Nd, respectively. M123 grains are
well oriented relative to each other, so that their $ab$-planes and c-axes are
almost parallel. The presence of Y211 or Nd422 grains, which can 
serve as pinning centers in the M123 matrix, leads to an increase in critical
current density. A melt-textured material usually has micro- and macrocracks
in the structure and pores (its density is 85-90 \% from the theoretical one).
The further increase of material density and improvement of mechanical
properties are of great importance.

\begin{figure}[b]
\begin{center}\leavevmode
\includegraphics[width=1\linewidth]{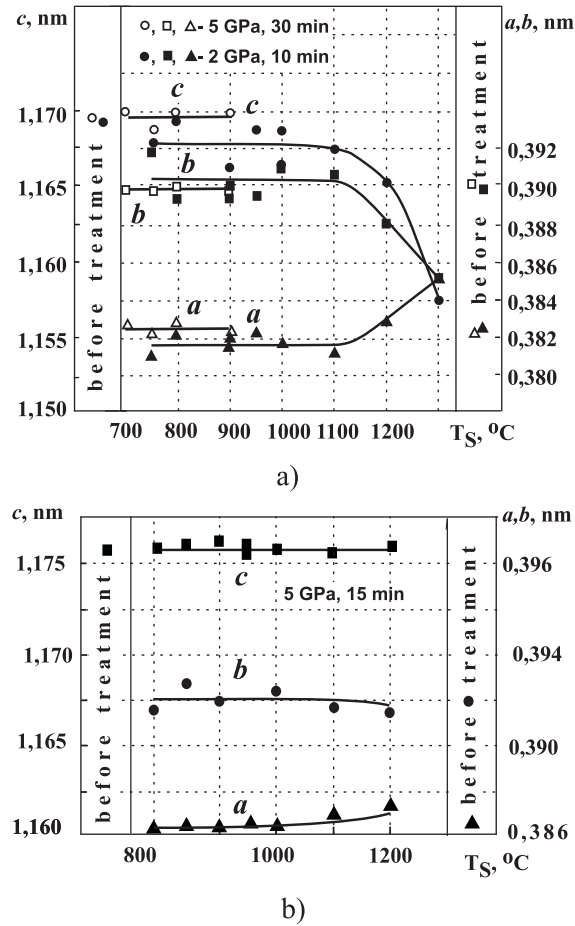}
\caption{
Crystallographic parameters $a, b, c$ versus treatment temperature ($T_s$)
before and after HP-HT treatment of
(a) YBa$_2$Cu$_3$O$_{7-\delta~}$-phase of MT-YBCO, treated at 2 and 5~GPa
(b) NdBa$_2$Cu$_3$O$_{7-\delta~}$-phase of MT-NdBCO, treated under
5~GPa}\label{fig1}\end{center}\end{figure}

\section{Method, results and discussion}
\begin{figure}[t]
\begin{center}\leavevmode
\includegraphics[width=1\linewidth]{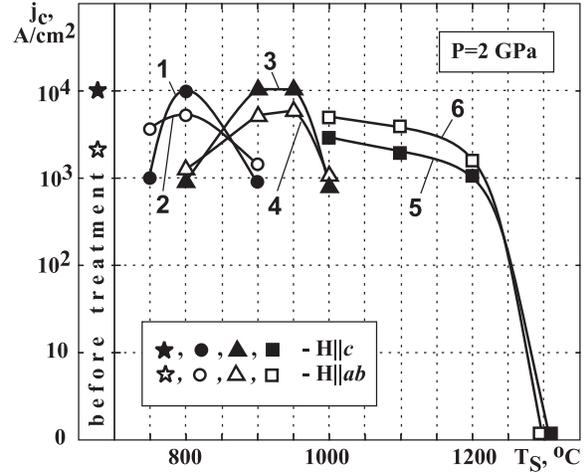}
\caption{
Schematic representation of the critical current density variations at
77 K  and zero magnetic field ($H$) of YBa$_2$Cu$_3$O$_{7-\delta~}$-phase
of MT-YBCO vs. the treatment
temperature ($T_s$) and time. Samples were treated 
at 2 GPa for 30 min (curves 1, 2), 15 min (3,4), and 10 min (5,6); solid dots
--- $H \| c$, open dots - $H \| ab$.
}\label{fig2}\end{center}\end{figure}

High pressure - high temperature conditions (HP-HT) have been created in
recessed-anvil-type HP apparatuses \cite{2}, where the treated sample has
been surrounded with precompacted monoclinic zircinia.
Fig.\ \ref{fig1} and \ref{fig2} present the variations of unit cell parameters and critical
current density ($j_c$) versus treatment conditions for MT-MBCO samples.
With an increase in treatment temperature the tetragonal Y123 structure forms
due to the increase of oxygen content up to more than 7 oxygen atoms per unit
cell  \cite{2} as witnessed by a decrease in the $c$-parameter. 
After optimal HP-HT treatment the $j_c$ in $ab$-plane of MT-MBCO (M = Y, Nd)
was unchanged: 10$^4$ A/cm$^2$, but some increase  in $j_c$ along $c$-axis have
been observed, which may be the result of densification and the increase in
dislocation density (for MT-YBCO from 10$^8$ up 
to 10$^{12}$ cm$^{-1}$ in (001) planes). Materials density increased from 5.7
to 6.3 g/cm$^3$ and from 5.5 to 
6.64 g/cm$^3$, 
Vickers microhardness from 3.4 to 5.3 GPa (4.91-N load) and
from 4.9 to 5.9 GPa (1.96-N load) for M = Y and Nd, respectively.

\end{document}